\documentstyle[preprint,tighten,aps,epsfig]{revtex}
\newcommand{\tr}{\mbox{Tr}\,}

\begin{document}
\title {\Large \bf The Ideal Mixing Departure in Vector Meson Physics}

\author {L.\ Epele, H.\ Fanchiotti and A.\ G.\ Grunfeld}
\address{\hfill \\Instituto de F\'{\i}sica La Plata CONICET, Dto.\ de
F\'{\i}sica, Fac.\ de Cs.\ Exactas, UNLP\\
C.C. 67, 1900 La Plata, Argentina} \maketitle
\begin{abstract}
In this work we study the departure for the ideal $\phi-\omega$
mixing angle in the frame of the Nambu-Jona-Lasinio model. We have
shown that in that context, the flavour symmetry breaking is
unable to produce the shifting in the mixing angle. We introduce a
nonet symmetry breaking in the neutral vector sector to regulate
the non-strange content of the $\phi$ meson. The phenomenon is
well reproduced by our proposal.
\end{abstract}

\vspace{1cm}

It is well known that it is not possible to use perturbative
expansions of QCD to describe low energy hadronic phenomena. For
that reason, a great deal of effective theories, preserving the
symmetries of QCD, have been developed to account for the main
properties of hadrons. One of the basic aspects common to both
schemes, is the chiral invariance of the strong interactions in
the massless limit. This symmetry is explicitly broken if quarks
are massive.

Chiral effective models \cite{leutw} have demonstrated to
reproduce the low energy hadron phenomena. The effective four
fermion Lagrangian, proposed by Nambu-Jona-Lasinio (NJL)
\cite{NJL}, is an intermediate step between QCD and effective
mesons theories. From NJL Lagrangian is it possible to obtain an
effective meson Lagrangian after proper bosonization \cite{Rein1}.

The NJL model is a profitable arena to study various phenomena
related to symetry brakdown in hadron physics. In particular, the
$\omega-\phi$ mixing was largely studied by theoretical and
experimental points of view since the begining of the sixties
\cite{Okubo}. Measurements show that the $\phi$ meson decays into
$\pi ^+ \pi^-$, violating isospin conservation and OZI rule
\cite{OZI,Oller}. In the present paper we will study the departure
of the $\omega-\phi$ ideal mixing angle in the NJL model, allowing
a non strange content in the $\phi$ meson. We will focus our
attention in the vector meson sector where the phenomenology we
are interested to describe takes place. It is important to note
that, in the framework of most theoretical models, the $\rho$ and
$\omega$ mesons are composed by $u$ and $d$ light quarks, whereas
the quark content of $\phi$ meson is purely strange, which do not
coincide with experimental data \cite{PDG}. Let us note that some
authors predict the departure from ideal mixing angle whithin
different approaches \cite{Bena,Spal,Jones}.

Our purpose is to explore the origin of the shifting in the
$\phi-\omega$ mixing angle in the frame of the NJL Lagrangian and
its connection with chiral symmetry. The goal is to find the
relation of that effect with the explicit breakdown of the $SU(3)$
flavour symmetry to $SU(2)$ symmetry with the assumption $m_u =
m_d \ne m_s$. The NJL model predicts the ideal mixing angle in the
vector sector in the process of diagonalization of the neutral
sector \cite{Volkov,Epele}.

In order to investigate the source of the departure of the ideal
mixing angle in the vector sector, we explore different scenarios.
First of all we present a brief summary of our previous results
\cite{Epele} where we have studied the explicit chiral symmetry
breaking in the NJL model when considering $m_u = m_d \ne m_s$. We
revisit that scheme focusing on the $\phi-\omega$ mixing angle.

As a second step we studied how QCD vertex corrections modify the
NJL coupling constant and its relation with the $\phi-\omega$
mixing angle. In the hadron scale we are dealing with, the non
perturbative gluon propagator can be approximated by an universal
constant leading to a local NJL Lagrangian \cite{Ebert}. The
chiral symmetry breakdown has no further effect in the tree level
approximation. However, the effective NJL coupling constant is
affected if we consider the QCD vertex corrections when chiral and
flavor symmetry breaking is considered. We model the contributions
emerging as a consequence of explicit breakdown of the SU(3)
flavor symmetry to SU(2) isospin symmetry, in four fermion
interactions coming from vertex corrections at QCD level. As a
consequence the symmetry breakdown becomes explicit in the
coupling constant of NJL Lagrangian. We will see in our present
work that, neither the explicit chiral symmetry breaking in the
NJL Lagrangian when considering quark masses, nor the
consideration of QCD vertex corrections, modify the ideal mixing
angle in the vector sector.

Then, as a final step, with the objective of investigating the
source of the departure of the ideal mixing angle for the vector
sector, we modify the coupling constant in the NJL Lagrangian
inspired by models for $\eta-\eta'$ physics \cite{thooft}. We add
a new parameter redefining the coupling constant in order to
separate the singlet state from the octet.

Let us start with the NJL Lagrangian \cite{NJL}
\begin{equation}
{\cal L} = \bar q (i \not\!\partial - \hat m_0)q + 2 G_1 \left[
(\bar q \frac{1}{2}\lambda^a q)^2 + (\bar q i \gamma_5
\frac{1}{2}\lambda^a q)^2 \right] - 2 G_2 \left[ (\bar q
\gamma^\mu \frac{1}{2}\lambda^a q)^2 + (\bar q \gamma^\mu
\gamma_5\frac{1}{2} \lambda^a q)^2 \right], \label{00}
\end{equation}
where $q$ denotes the $N$-flavour quark spinor, $\lambda^a$,
$a=0,\dots,N^2-1$ are the generators of the $U(N)$ flavor group
(we normalize $\lambda^0=\sqrt{2/N}\,${\bf 1}) and $\hat m_0$
stands for the current quark mass matrix. The coupling constants
$G_1$ and $G_2$, as well as the quark masses, are introduced as
free parameters of the model. In the absence of the mass term, the
NJL Lagrangian shows at the quantum level the $SU(N)_A \otimes
SU(N)_V \otimes U(1)_V$ symmetry characteristic of massless QCD.

It is possible to reduce the fermionic degrees of freedom to
bosonic ones by bosonization technique. By means of the
Stratonovich identity, the vector--vector coupling in (\ref{00})
can be transformed as
\begin{equation}
- 2 G_2 \left(\bar{q} \gamma^\mu \frac{1}{2}\lambda^a q \right)^2
\rightarrow -\frac{1}{4 G_2} \tr V_\mu^2 + i \bar{q}\gamma^\mu
V_\mu q, \label{01}
\end{equation}
where
\begin{equation}
V_\mu \equiv -i \displaystyle \sum_{a=0}^8 V_\mu^a\;\lambda^a/2.
\label{01a}
\end{equation}
The spin 1 fields $V_\mu^a$ can be identified with the usual nonet
of vector mesons as in \cite{Epele}, which transform in such a way
to preserve the chiral symmetry of the original NJL Lagrangian
(and therefore that of QCD). Notice that the first term in the
right hand side of Eq.\ (\ref{01}) is nothing but a mass term for
the vector fields $V_\mu^a$, thus the vector--meson masses are
governed by the coupling $G_2$ in the NJL Lagrangian. It can be
seen that these masses are degenerate in the limit where the quark
masses are degenerate. The quark fields can be integrated out,
leading to an effective Lagrangian which only contains bosonic
degrees of freedom. This procedure can be carried out by taking
into account the generating functional and performing the
calculation of the fermion determinant (a detailed analysis can be
found in \cite{Rein1}). A similar procedure can be followed for
the full NJL Lagrangian (\ref{00}), leading to the interactions
involving scalar, pseudoscalar and axial--vector bosons. In this
way, the final effective Lagrangian is written only in terms of
spin 0 and spin 1 colorless hadron fields.

In our previous work \cite{Epele} we have studied the explicit
chiral symmetry breaking in NJL model when $m_u = m_d \ne m_s$,
obtaining the ideal mixing in the process of diagonalizing the
neutral vector meson sector. We have performed the bosonization by
carrying out an expansion of the fermion determinant, which gives
rise to a set of one--loop Feynman diagrams \cite{Eguchi}. The
generating functional gives rise to effective kinetic terms for
the spin-1 vector mesons via one--loop diagrams, giving the
following contribution to the vector meson self-energy
\begin{equation}
i N_c \int \frac{d^4 k}{(2\pi)^4} \tr \frac{\not\!k - \not\!p +
m_1}{(k-p)^2 - m_1^2} \lambda^a \gamma_\mu \frac{\not\!k +
m_2}{k^2 - m_2^2}\lambda^b \gamma_\nu\;, \label{pimunu}
\end{equation}
where $m_1$ and $m_2$ are the constituent masses of the quarks
entering the loop, $N_c$ is the number of colors, and the trace
acts over the flavor and Dirac indices.

We will take only the leading order in the external momentum $p$,
which means to evaluate the integral at $p=0$ after extracting the
relevant kinematical factors. In this case, this is equivalent to
consider only the divergent piece of (\ref{pimunu}):
\begin{equation}
\Pi_{\mu\nu}^{(V)} = I_2(m_1, m_2)\, \left[\frac{1}{3}\,(p_\mu
p_\nu - p^2 g_{\mu\nu}) + \frac{1}{2}\, (m_2 - m_1)^2\right]\,,
\label{pimunudiv}
\end{equation}
where
\begin{equation}
I_2(m_i, m_j) \equiv -i\,\frac{N_c}{(2 \pi)^4} \int d^4k
\frac{1}{(k^2 - m_i^2)(k^2 - m_j^2)}\,.
\end{equation}
In order to regularize the divergence we use the proper--time
regularization scheme with a cut--off $\Lambda$, which will be
treated as a free parameter of the model. We obtain
\begin{equation}
I_2(m_i, m_j) = \frac{N_c}{16 \pi^2}\int_0^1 dx\;
\Gamma\left(0,\frac{(m_i^2 - m_j^2) x + m_j^2
}{\Lambda^2}\right)\,. \label{idos}
\end{equation}

{}From (\ref{pimunudiv}), the kinetic terms for the vector mesons
in the effective Lagrangian are given by
\begin{eqnarray}
{\cal L}_{kin}^{(V)} & = & -\frac{1}{4}\,\frac{2}{3}\, I_2(m_u,
m_u) \left[\rho_{\mu\nu} \rho^{\mu\nu} + 2 \rho^+_{\mu\nu}
{\rho^-}^{\mu\nu} + \omega_{\mu\nu} \omega^{\mu\nu} + \alpha\,
\phi_{\mu\nu} \phi^{\mu\nu} \nonumber \right.
\\ & & \left. + 2 \beta \left( {K^\ast}^+_{\mu\nu} {K^\ast}^{-\mu\nu} +
{K^\ast}^0_{\mu\nu} \bar K^{\ast 0\mu\nu} \right)\right] \, ,
\label{lkin}
\end{eqnarray}
where $V^{\mu\nu} \equiv \partial^\mu V^\nu - \partial^\nu V^\mu$,
and
\begin{equation}
\alpha = \frac{I_2(m_s, m_s)}{I_2(m_u, m_u)} \;, \qquad\qquad
\beta = \frac{I_2(m_u, m_s)}{I_2(m_u, m_u)}
\end{equation}
parameterize the magnitude of the chiral symmetry breaking.

The kinetic Lagrangian in (\ref{lkin}) has been expressed in terms
of the vector fields, with the additional rotation
\begin{eqnarray}
\omega_8 & = & \phi \cos\theta_0 + \omega \sin\theta_0 \nonumber\\
\omega_1 & = & -\phi \sin\theta_0 + \omega \cos\theta_0
\label{rot}
\end{eqnarray}
which diagonalizes the neutral sector. It is easy to see that in
this model the rotation is ``ideal'', i.e., the spin 1 mass
eigenstates $\rho$ and $\omega$ are composed by pure light $u$ and
$d$ quarks, while the $\phi$ meson is a bound state $\bar s s$.
The ideal rotation angle is given by the well known relation
$\sin\theta_0 = 1/\sqrt{3}$.

The mass terms for the vector mesons are given by the $(V_\mu)^2$
term in (\ref{01}), plus a divergent one--loop contribution given
by the second term in the square brackets in (\ref{pimunudiv}),
which vanishes in the chiral limit. This leads to
\begin{eqnarray}
{\cal L}_{mass}^{(V)} & = & \frac{1}{8 G_2}\left[\rho_\mu \rho^\mu
+ 2 \rho^+_\mu {\rho^-}^\mu + \omega_\mu \omega^\mu + \phi_\mu
\phi^\mu + 2 {K^\ast}^+_\mu {K^\ast}^{-\mu} + 2 {K^\ast}^0_\mu
\bar K^{\ast 0\mu}\right] \nonumber \\ & & + (m_s - m_u)^2 \beta\,
I_2(m_u,m_u) ({K^\ast}^+_\mu {K^\ast}^{-\mu} + {K^\ast}^0_\mu \bar
K^{\ast 0\mu})\;. \label{lvmass}
\end{eqnarray}
Notice that (as expected) the mass terms turn out to be diagonal
in the ($\omega,\phi$) basis.

We proceed now to the wave function renormalization required by
the kinetic terms in (\ref{lkin}). The vector meson fields can be
properly redefined by $V_\mu \to Z^{1/2}_V\, V_\mu$, with
\begin{eqnarray}
Z_\rho^{-1} & = &  Z_\omega^{-1}  = \frac{2}{3}\,I_2(m_u,m_u) \label{vrena} \nonumber\\
Z_{K^\ast}^{-1} & = &  \frac{2}{3}\,\beta \,I_2(m_u,m_u) = \beta\, Z_\rho^{-1} \nonumber\\
Z_\phi^{-1} & = & \frac{2}{3}\,\alpha\,I_2(m_u,m_u) = \alpha\, Z_\rho^{-1} \;.
\label{vren}
\end{eqnarray}
Then from (\ref{lvmass}) one obtains
\begin{equation}
m_\rho^2 = m_\omega^2 = \frac{Z_\rho}{4 G_2}\;,\qquad m_{K^*}^2 =
\frac{m_\rho^2}{\beta} + \frac{3}{2}(m_s - m_u)^2\;,\qquad
m_\phi^2 = \frac{m_\rho^2}{\alpha}\;, \label{vmass}
\end{equation}
thus the $\phi$ meson mass can be written in terms of the $\rho$
mass and the chiral symmetry breaking parameter $\alpha$. In the
case of the $K^\ast$, the corresponding mass relation includes
both the parameter $\beta$ and a quark--mass dependent
contribution that arises from the loop.

From these previous results, the explicit chiral symmetry breaking
in the NJL model taking into account the explicit breakdown of the
$SU(3)$ flavour symmetry to $SU(2)$ isospin symmetry, do not leads
to a departure of the ideal mixing angle in the vector meson
sector.

As we mentioned before, another mechanism should be responsible
for such effect. Our aim is to perform a further analysis,
focusing our attention to the four-fermion interaction in the NJL
model. We will study how four--fermion interaction is modified
when considering the flavour symmetry breaking and its
consequences on the vector mixing angle. The effect of chiral
symmetry breakdown, when considering quark mass terms, do not
modify the coupling. However, one can expect that the explicit
flavour symmetry breaking at QCD level should manifest in the
quarks couplings in the NJL model through the vertex corrections.

In QCD, vertex corrections depend on quark propagators and
consequently, on quarks masses, as shown in Fig.\ 1. If quarks
masses are degenerated, the QCD vertex contributions are the same
for all couplings. Nevertheless, if we consider the case of
explicit breakdown of the $SU(3)$ flavor symmetry to $SU(2)$
isospin symmetry, the internal lines in Fig.\ 1 will have
different propagators due to different quarks masses. As a
consequence, the contributions coming from vertex corrections at
QCD level should modify the four fermion interaction in the NJL
Lagrangian. We modelled that effect introducing in the NJL
Lagrangian as new chiral symmetry breaking parameter in the
strange current.

Let us start writing the terms containing vector current-current
interaction in (\ref{00}). As we are interested in the departure
of the $\phi-\omega$ ideal mixing angle, we will only concentrate
in the neutral vector current--current terms; the extension to the
full Lagrangian is straightforward
\begin{equation}
-2 G_2 \left(\bar{q}\; \gamma_\mu \;\frac{\lambda_a}{2} q \right)^2 =
-
2 G_2\left[\left(\bar{u}\;\bar{d}\;\bar{s}\right)\;\gamma_\mu
\;\frac{\lambda_a}{2} \left (
\begin{array}{c}  u \\  d \\ s
\end{array} \right)\right]^2\;;
\end{equation}
as we are considering three flavors (u, d, s), $\lambda_a$ are the Gell-Mann
matrices with the normalization $\lambda^0=\sqrt{2/N}\,${\bf 1}. Let us rewrite the
diagonal terms, i.e. with $\lambda_0, ~\lambda_3,~\lambda_8$,
\begin{eqnarray}
J_0&=&\frac{\sqrt{2}}{2\sqrt{3}}{\bar{u}}\gamma_\mu u +
\frac{\sqrt{2}}{2\sqrt{3}}{\bar{d}}\gamma_\mu d +
\frac{\sqrt{2}}{2\sqrt{3}} {\bar{s}}\gamma_\mu s \nonumber\\
J_3 &=& \frac{1}{2}{\bar{u}}\gamma_\mu u -
\frac{1}{2}{\bar{d}}\gamma_\mu d  \nonumber\\ J_8
&=&\frac{1}{2\sqrt{3}}{\bar{u}}\gamma_\mu u +
\frac{1}{2\sqrt{3}}{\bar{d}}\gamma_\mu d -
\frac{2}{2\sqrt{3}}{\bar{s}}\gamma_\mu s,
\end{eqnarray}
here we have set $J_a = {\bar{q}} \gamma_\mu (\lambda_a/2) q$ (for
simplicity we have omitted the index $\mu$). Defining $U =
\bar{u}\gamma_\mu u,~D = \bar{d}\gamma_\mu d$ and $ S =
\bar{s}\gamma_\mu s$, it is easy to see that when $m_u = m_d =
m_s$, the sum of the these three terms (squared) presents flavor
symmetry $SU(3)$ and it is proportional to $(U^2 + D^2 + S^2)$.

We will focus on the case when $m_u = m_d \ne m_s$. We choose the
$S$ current--current interaction in the NJL Lagrangian in the
following way
\begin{equation}
S^2 \rightarrow (1+\epsilon)^2 S^2 \label{03}
\end{equation}
or equivalently
\begin{equation}
s \rightarrow (1+\epsilon) s, \label{04}
\end{equation}
where the $\epsilon$ parameter puts in evidence the contributions
coming from QCD vertex corrections when considering the explicit
breakdown of flavor symmetry. Then, taking into account (\ref{03})
\begin{equation}
\left(U^2 + D^2 + S^2 \right) \to \left( U^2 + D^2 + S^2 \right) +
\epsilon (2+\epsilon) S^2.
\end{equation}
We rewrite the diagonal current--current interactions at ${\cal
{O}}(\epsilon^2)$
\begin{eqnarray}
J_0&=&\frac{\sqrt{2}}{2\sqrt{3}} (U + D + S) \to J'_0 =
\frac{\sqrt{2}}{2\sqrt{3}} (U + D + S + \epsilon S) = (1 +
\frac{\epsilon}{3})J_0 - \frac{\sqrt{2}}{3}\epsilon J_8
\nonumber\\ J_3&=&(U - D) \nonumber\\
J_8&=&\frac{\sqrt{2}}{2\sqrt{3}} (U + D - 2S) \rightarrow J'_8 =
\frac{\sqrt{2}}{2\sqrt{3}} (U + D - 2S - 2\epsilon S) =(1+
\frac{2\epsilon}{3}) J_8 - \frac{\sqrt{2}}{3}\epsilon J_0,
\end{eqnarray}
where the primed currents are expressed in terms of a mixture of
the non primed ones, regulated by $\epsilon$ parameter.

The NJL Lagrangian vector terms can be expressed in terms of the
primed currents, containing the $\epsilon$ dependence, as follows
\begin{equation}
{\cal{L}'}^{(V)}_{NJL} = \bar{q}(i \not\!\partial - \hat{m}_0)q -
2 G_2 \sum_{a=0}^{8} J'^2_a.
\end{equation}
In this case, following the procedure presented in \cite{Epele},
the quark fields can be integrated out, leading to an effective
Lagrangian which only contains bosonic degrees of freedom. We have
performed the bosonization by carrying out an expansion of the
fermion determinant, which gives rise to a set of 1-loop Feynman
diagrams. Taking into account (\ref{01}), the final Lagrangian can
be written in terms of the bosonic fields
\begin{eqnarray}
\omega_1 &=& \left(1+\frac{\epsilon}{3}\right) \omega'_1
 - \frac{\sqrt{2}}{3} \epsilon \; \omega'_8 \nonumber\\
\omega_8 &=& \left(1 + \frac{2 \epsilon}{3}\right) \omega'_8
 - \frac{\sqrt{2}}{3} \epsilon \; \omega'_1,
\label{primed}
\end{eqnarray}
warranting that kinetic energy has the same expression as in
(\ref{lkin}), with no dependence in $\epsilon$ parameter. Note
that, from (\ref{01a}), identifying $V_\mu^a$ as the usual nonet
of vector mesons \cite{Epele}, $V_0$ and $V_8$ correspond to
$\omega_1$ and $\omega_8$ respectively.

Then, keeping the first order in the $\epsilon$ power expansion,
the neutral vector mass terms are
\begin{equation}
{\cal{L}}_{mass}^{(V)} = \frac{1}{8 G_2}\left[(1 -
\frac{2\epsilon}{3})\omega_1^2 + \rho^2 + (1 -
\frac{4\epsilon}{3})\omega_8^2 + \frac{4\sqrt{2}}{3} \epsilon \;
\omega_1 \omega_8 \right]. \label{mass0}
\end{equation}

Both kinetic and mass terms are non diagonal in the neutral
sector. In order to diagonalize them simultaneously it is
necessary to introduce the following rotation which includes two
different angles
\begin{eqnarray} \omega_8 & = &
\phi \cos\theta_1 + \omega \sin\theta_2 \nonumber\\ \omega_1 & = &
-\phi \sin\theta_1 + \omega \cos\theta_2 \;. \label{rota}
\end{eqnarray}
In the literature, some authors \cite{Bena,Spal,Jones} obtain by
different procedures two different mixing angles. Replacing
(\ref{rota}) in (\ref{mass0}) we obtain the following form
\begin{equation}
-2 \left(1 - \frac{2\epsilon}{3}\right)\sin \theta_1\cos \theta_2
+ 2 \left(1 - \frac{4\epsilon}{3}\right)\sin \theta_2\cos \theta_1
+ \frac{4 \sqrt{2}}{3} \epsilon \;(\cos \theta_1\cos \theta_2 -
\sin \theta_1\sin \theta_2) = 0. \label{diag1}
\end{equation}
which is the mass terms diagonalization condition.

The above expression is again satisfied with the ideal mixing
angle as well as the kinetic terms. As a consequence, the non
strange content of $\phi$ meson does not arise from the coupling
constant modification when considering QCD vertex corrections with
$m_u = m_d \ne m_s$.

In order to estimate the magnitude of the $\epsilon$, we express
the vector meson masses in terms of that parameter. We proceed to
the wave function renormalization required by kinetic terms of
(\ref{lkin}), redefining the meson fields as in (\ref{vren}).
Considering the mass contributions coming from the divergent
1-loop contribution \cite{Epele}, the vector meson masses,
expressed in terms of the chiral symmetry breaking parameters are
\begin{equation}
m_\rho^2 = m_\omega^2 = \frac{Z_\rho}{4 G_2},\;\;\; m_{K^*}^2 = \frac{m_\rho^2
(1-\epsilon)}{\beta} + \frac{3}{2}(m_s - m_u)^2,\;\;\; m_\phi^2 =
\frac{m_\rho^2(1-2\epsilon)}{\alpha}\;. \label{vmasse}
\end{equation}

Taking into account the experimental value for the $\phi$ mass
\cite{PDG}, we can estimate the value for $\epsilon$. In our
calculation we have supposed that the chiral symmetry breaking
parameters $\alpha$ and $\beta$, are not modified by considering
the vertex corrections at QCD level (here we use the
phenomenological values for these parameters obtained in
\cite{Epele}). In this way, we estimate the value for the
$\epsilon$ parameter
\begin{equation}
\epsilon \simeq -0.03.
\end{equation}
Therefore, the coupling constant in the neutral sector (those
terms in NJL Lagrangian with $S$ current-current interactions) and
in the charged vector sector are $0.94 G_2$ and $0.97 G_2$
respectively. However, those tiny but non-vanishing vertex
corrections that modify meson masses, are not able to shift the
ideal mixing angle.

Our results lead us to conclude that the mechanism responsible for
the ideal mixing departure has no source neither in the explicit
chiral symmetry breaking in the NJL Lagrangian \cite{Epele} when
considering quark masses throughout QCD vertex corrections. The
inclusion of chiral symmetry breaking parameters $\alpha$ and
$\beta$ in \cite{Epele}, as well as the parameter which takes into
account the QCD vertex corrections $\epsilon$, do not lead to any
change in the mixing angle in the vector sector. That means that
another mechanism will be responsible to allow a non vanishing $u,
d$ content of meson $\phi$ to permit decays as $\phi \to \pi^+
\pi^-$ \cite{PDG}. In our opinion, inspired by the phenomenology
of pseudoscalar sector, one possible approach to solve that
puzzle, is considering a new parameter. In the pseudoscalar
sector, the presence of the $U(1)$ anomaly breaks the $U(3)$
symmetry down to $SU(3)$, leading to the mass splitting between
the observed $\eta$ and $\eta$' physical states. In the NJL model
this effect is included throughout the 't Hooft interaction
\cite{thooft}. Another way to take into account the anomaly is
introducing the $\eta-\eta$' mixing angle as a parameter of the
model. Inspired in this peculiar physics, we have tested the
sensibility of mixing angle in the vector sector including a
parameter $\delta$ to force a nonet symmetry breaking. We proceed
as follows
\begin{equation}
- 2 G_2 {J'}_0^2 \to - 2 G'_2 {J'}_0^2
\end{equation}
with $ G'_2 \equiv \delta G_2$. This new parameter, after proper
bosonization, can be absorbed in the quadratic term as follows
\begin{equation}
-2 {G'}_2 {J'}_0^2 \to \;\left(\frac{-1}{4 {G'}_2} {V'}_0^2 +
{V'}_0 {J'}_0 \right).
\end{equation}
We have kept the $\epsilon$ dependence in mass terms and we will
see the consequences in our calculations. Then proceeding as
before, let us rewrite the neutral mass terms including the new
parameter $\delta$
\begin{eqnarray}
{\cal{L}}_{mass}^{(V)} = \frac{1}{8 G_2}\left[\frac{1}{\delta}
\left(1 - \frac{2\epsilon}{3}\right) \omega_1^2 + \rho^2 + \left(1
- \frac{4\epsilon}{3}\right)\omega_8^2 + \frac{2\sqrt{2}}{3}
\epsilon \;\left(1+ \frac{1}{\delta}\right)\; \omega_1 \omega_8
\right]. \label{mass1}
\end{eqnarray}
Note that the mixing in quadratic neutral sector term is regulated
by the $\delta$ and $\epsilon$ parameters. Is easy to see that in
the limit $\delta \to 1$ we reobtain (\ref{mass0}). As before, we
have performed the following rotation to diagonalize both kinetic
and quadratic terms, considering two different angles
\begin{eqnarray}
\omega_8 &=& \phi\;\cos \theta_1 + \omega\;\sin \theta_2 \nonumber\\
\omega_1 &=& - \phi\;\sin \theta_1 + \omega \;\cos \theta_2.
\label{angulos}
\end{eqnarray}
We found that kinetic terms of (\ref{lkin}) are diagonal when the
following condition is satisfied
\begin{equation}
(1+2 \alpha) \tan \theta_2 - (2 + \alpha) \tan \theta_1 + \sqrt{2}
(1 - \alpha)(1 - \tan\theta_1 \; \tan\theta_2)\;=\;0,
\label{diagkin}
\end{equation}
and mass terms are diagonal if
\begin{eqnarray}
\frac{-2}{\delta}\left(1 - \frac{2\epsilon}{3}\right)\sin
\theta_1\cos \theta_2 + 2 \left(1 - \frac{4\epsilon}{3}\right)\sin
\theta_2\cos \theta_1 +\nonumber\\
+ \frac{2\sqrt{2}}{3} \epsilon \;\left(1+ \frac{1}{\delta}\right)
(\cos \theta_1\cos \theta_2 - \sin \theta_1\sin \theta_2) = 0\;.
\label{diagmass}
\end{eqnarray}
It is straightforward to see that the ideal mixing angle does not
satisfy the above conditions simultaneously. The above relations
become trivial when $m_u = m_d = m_s$, then $\alpha = 1$ and
$\epsilon = 0$ are excluded in the following calculations. If
$\delta=1$ is considered, we reobtain (\ref{diag1}).

We proceed now to obtain the vector meson masses in terms of
$\delta$ and the two mixing angles. Our intention is computing the
magnitude of those parameters. Considering the wave function
renormalization (\ref{vren}) as before, from (\ref{mass1}) we
obtain
\begin{eqnarray}
m_\rho^2 &=& \frac{Z_\rho}{4 G_2},\;\;\; m_{K^*}^2 =
\frac{m_\rho^2
(1-\epsilon)}{\beta} + \frac{3}{2}(m_s - m_u)^2\nonumber\\
m_\omega^2 &=& m_\rho^2 \left[\frac{1}{\delta}\left(1 -
\frac{2\epsilon}{3}\right)\cos^2 \theta_2 + \left(1 -
\frac{4\epsilon}{3}\right)\sin^2 \theta_2  + \frac{2 \sqrt{2}}{3}
\epsilon \left(1 + \frac{1}{\delta} \right)
\sin\theta_2 \cos\theta_2 \right] \nonumber\\
m_\phi^2 &=&
\frac{m_\rho^2}{\alpha}\;\left[\frac{1}{\delta}\left(1 -
\frac{2\epsilon}{3}\right)\sin^2 \theta_1 + \left(1 -
\frac{4\epsilon}{3}\right)\cos^2 \theta_1 - \frac{2 \sqrt{2}}{3}
\epsilon \left(1 + \frac{1}{\delta} \right) \sin\theta_1
\cos\theta_1\right]. \label{deltamass}
\end{eqnarray}

Experimentally, the mixing angle is near 35$^\circ$ \cite{PDG}
(-0,3$^\circ$ apart from the ideal mixing angle), then we choose
\begin{equation}
\tan \theta_1\;=\;\frac{1}{\sqrt{2}} \;+\;x \label{t1}
\end{equation}
as the shifting in the ideal mixing angle. Replacing (\ref{t1}) in
condition (\ref{diagkin}), we obtained
\begin{equation}
\tan \theta_2\;=\;\frac{1}{\sqrt{2}} \;+\;\frac{x}{\alpha}\;+\;
{\cal{O}}(x^2). \label{t2}
\end{equation}
We have replaced both $\tan \theta_1$ and $\tan \theta_2$ in
condition (\ref{diagmass}), together with the definition $d =
(1/\delta) - 1$. Then, at ${\cal{O}}(d^2)$ and neglecting terms
containing $d \epsilon$, the expression for $x$ is the following
\begin{equation}
x\;\simeq -\frac{\sqrt{2}\; d}{2\; (1 - \frac{1}{\alpha})}.
\label{x}
\end{equation}

It is interesting to express the physical states $\phi_F$ and
$\omega_F$ in terms of the ``ideal ones'' ($\phi_I$ and
$\omega_I$) when considering a shifting in the ideal mixing angle
\begin{eqnarray}
\phi_F &=& \phi_I\ - \frac{1}{\alpha} \; x \; \omega_I  \nonumber\\
\omega_F &=& \omega_I + x \; \phi_I. \label{fisicos}
\end{eqnarray}
From the above expressions we can see that the non strange decays
of the $\phi$ meson is controlled by $x$, i. e. the shifting in
$\theta_1$ mixing angle. Then, from (\ref{t1}) and (\ref{x})
together with the experimental value for the shifting in the ideal
mixing angle, we determined the phenomenological value for the
parameter $\delta$, that leads to the following relation between
the coupling constant
\begin{equation}
G'_2/G_2 \simeq 1.005.
\end{equation}
Though tiny, this difference in the coupling constant becomes
crucial to model the departure from the ideal mixing angle.

Summing up, we have devoted our phenomenological analysis to study
the departure from the ideal mixing angle in the vector meson
sector in the frame of the NJL model, considering the flavour
symmetry breaking when $m_u = m_d \ne m_s$. We have analyzed
different mechanisms separately and together.

As a starting point we have revisit our previous results
\cite{Epele} focusing on the $\phi-\omega$ mixing angle. We show
that the explicit chiral symmetry breaking when considering quark
masses in NJL Lagrangian, does not lead to a non strange content
of $\phi$ meson.

As a second step, to explore another possible source of the ideal
mixing departure, we have include in our analysis the
phenomenology associated with vertex corrections at QCD level
which accounts of flavour symmetry breaking. As a consequence, the
effective couplings in NJL Lagrangian containing strange quarks in
the currents, are modified throughout the parameter $\epsilon$.
After bosonization, both kinetic and mass terms are diagonalized
again with the ideal mixing angle. Consequently, in this
framework, the $\phi$ meson is still composed by $s \bar{s}$
quarks, then, another mechanism should be responsible for the
ideal-mixing departure. Throughout the expressions for the vector
meson masses, we have estimated the phenomenological value for the
parameter $\epsilon$.

As a final step, inspired in the peculiar physics of the neutral
pseudoscalar sector, we have forced a nonet symmetry breaking in
the vector sector.  For that purpose we have included the $\delta$
parameter to test the sensibility of the $\phi-\omega$ mixing
angle. In this scheme, after proper bosonization, we have
diagonalized simultaneously both kinetic and mass terms with two
different non ideal angles, allowing a non strange content of
$\phi$ meson. We have expressed de physical $\phi$ and $\omega$
mesons, in terms of the shifting in the ideal mixing. From the
expressions (\ref{fisicos}) it can be seen that only one of the
mixing angles is related with physical observables. We have
obtained a phenomenologycal value for the parameter $\delta$, in
terms of the shifting in the ideal mixing angle, $\alpha$ and
$\epsilon$ parameters.

\acknowledgments

A. G. Grunfeld acknowledges CLAF-CNPq for financial support. This
research was partially supported by ANPCyT (Argentina).

\begin{figure}[ht]
    \begin{center}
       \setlength{\unitlength}{1truecm}
       \leavevmode
       \hbox{
       \epsfysize=4.5cm
       \epsffile{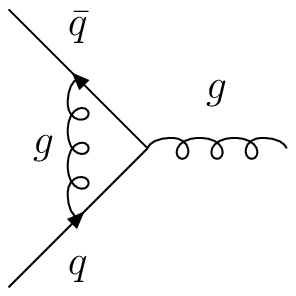}
       }
    \end{center}
    \caption{\footnotesize{One--loop correction to quark-gluon
    vertex in QCD.}}
    \protect\label{fig2}
\end{figure}
%

\end{document}